\documentclass[namedreferences]{solarphysics}
\usepackage[optionalrh,solaenum]{spr-sola-addons} 
\usepackage{graphicx}        
\usepackage{color}           
\usepackage{url}             
\usepackage{solaheader}
\newcommand{\solareceiveddate}{4 April 2011}
\newcommand{\solaaccepteddate}{11 November 2011}
\newcommand{\Xdoi}{s11207-011-9907-2}

\def\gsim{\,\lower3pt\hbox{$\sim$}\llap{\raise2pt\hbox{$>$}}\,}
\def\lsim{\,\lower3pt\hbox{$\sim$}\llap{\raise2pt\hbox{$<$}}\,}
\newcommand{\be}{\begin{equation}}
\newcommand{\ee}{\end{equation}}
\newcommand{\bex}{\begin{equation}\notag}
\newcommand{\eex}{\end{equation}\notag}
\newcommand{\bea}{\begin{eqnarray}}
\newcommand{\eea}{\end{eqnarray}}
\newcommand{\beax}{\begin{eqnarray*}}
\newcommand{\eeax}{\end{eqnarray*}}
\newcommand{\ba}{\begin{array}}
\newcommand{\ea}{\end{array}}

\newcommand{\grad}{ {\bf \nabla } }

\newcommand{\vecB}{{\mathbf B}}

\newcommand{\vecF}{{\mathbf F}}

\newcommand{\vecT}{{\mathbf T}}

\newcommand{\vecf}{{\mathbf f}}
\newcommand{\vecv}{{\mathbf v}}

\newcommand{\vecnhat}{{\mathbf {\hat n}}}
\newcommand{\vecrhat}{{\mathbf {\hat r}}}

\newcommand{\apj}{    {\it Astrophys. J.}}
\newcommand{\apjl}{    {\it Astrophys. J. Lett.}}
\newcommand{\apjs}{    {\it Astrophys. J. Suppl.}}

\newcommand{\jgr}{    {\it J. Geophys. Res.}}
\newcommand{\mnras}{  {\it Mon. Not. Roy. Astron. Soc.}}
\newcommand{\nat}{    {\it Nature}}

\newcommand{\pasj}{   {\it Pub. Astron. Soc. Japan}}

\newcommand{\solphys}{{\it Solar Phys.}}

\begin{document}

\begin{article}

\begin{opening}

\title{Global Forces in Eruptive Solar Flares:  The
Lorentz Force Acting on the Solar Atmosphere and the Solar Interior}

\author{G.H.~\surname{Fisher}$^{1}$\sep
        D.J.~\surname{Bercik}$^{1}$\sep
        B.T.~\surname{Welsch}$^{1}$\sep
        H.S.~\surname{Hudson}$^{1}$      
       }
\runningauthor{G.H. Fisher \it{et al.}}
\runningtitle{Global Lorentz Force}

   \institute{$^{1}$ Space Sciences Laborary, University of California, Berkeley, CA, USA\\
                     email: \url{fisher@ssl.berkeley.edu}, \url{bercik@ssl.berkeley.edu}, \url{welsch@ssl.berkeley.edu}, \url{hhudson@ssl.berkeley.edu}
             }

\begin{abstract}
We compute the change in the Lorentz force integrated over the outer
solar atmosphere implied by observed changes in vector
magnetograms that occur during large, eruptive solar flares.  
This force perturbation should be balanced
by an equal and opposite force perturbation 
acting on the solar photosphere and solar
interior.  The resulting expression for the estimated
force change in the solar interior generalizes the earlier expression presented
by Hudson, Fisher, and Welsch (\textit{Astron. Soc. Pac.}, \textbf{CS-383}, 221, 2008), 
providing horizontal as well as
vertical force
components, and provides a more accurate result for the vertical
component of the perturbed force.
We show that magnetic eruptions should result in the magnetic
field at the photosphere becoming more horizontal, and hence should result in
a downward (towards the solar interior)
force change acting on the photosphere and solar interior, 
as recently argued from an analysis of
magnetogram data by Wang and Liu (\textit{Astrophys. J. Lett.} \textbf{716}, L195, 2010).  
We suggest the existence of
an observational relationship between the force change computed
from changes in the vector magnetograms, 
the outward momentum carried by the ejecta from the flare, and
the properties of the helioseismic disturbance driven by 
the downward force change.  We use the impulse driven by the 
Lorentz-force change in the outer solar atmosphere to derive an 
upper limit to the
mass of erupting plasma that can escape from the Sun.
Finally, we compare the expected Lorentz-force change at the photosphere
with simple estimates from flare-driven gasdynamic disturbances and from
an estimate of the perturbed pressure from radiative backwarming of the
photosphere in flaring conditions.
\end{abstract}
\keywords{Active Regions, Magnetic; Coronal Mass Ejections, Theory; 
Flares, Dynamics; Flares, Relation to Magnetic Field; 
Helioseismology, Theory; Magnetic fields, Corona}
\end{opening}

\section{Introduction}
\label{Introduction} 

Eruptive flares and CMEs result from global reconfigurations of the 
magnetic-field 
in the solar atmosphere.  Recently, signatures of this magnetic field
change have been detected in both vector and line-of-sight magnetograms, 
the maps of the vector and the line-of-sight component of 
the photospheric magnetic field, respectively. 
Is there a relationship between
this measured field change and properties of the eruptive phenomenon?  What
is the relationship between forces acting on the outer solar atmosphere
and those acting on the photosphere and below, in the solar convection zone?

We will attempt to address these questions by considering the action of
the Lorentz force over a large volume in the solar atmosphere 
that is consistent with observed changes in the
photospheric magnetic field, and we will discuss how one can derive
observationally testable limits on eruptive flare or 
CME mass that are based on these
force estimates.  We will also provide more context for the recent
result of \inlinecite{Hudson2008}, who present an estimate for the inward force
on the solar interior driven by changes observed in magnetograms.  In addition,
we will provide additional interpretation of the recent observational 
results of \inlinecite{Wang2010}, who find from vector-magnetogram
observations that
the force acting on the photosphere and interior is nearly always downward,
and \inlinecite{Petrie2010}, who find similar results from a statistical
study using line-of-sight magnetograms.

Finally, we will compare the downward impulse from changes in the Lorentz
force with pressure impulses from heating by
energetic-particle release during flares,
and with radiative backwarming during flares, with the goal of describing
what future work is necessary to assess which physical
mechanisms produce the largest change in force density at the photosphere,
and hence which might be most effective in driving 
helioseismic waves ($e.g.$ \opencite{Kosovichev2011}) into the solar interior.

\section{The Lorentz Force Acting on the Upper Solar Atmosphere}
\label{section:maxwell}

\begin{figure}    
   \centerline{\includegraphics[width=1.0\textwidth,clip=]
   {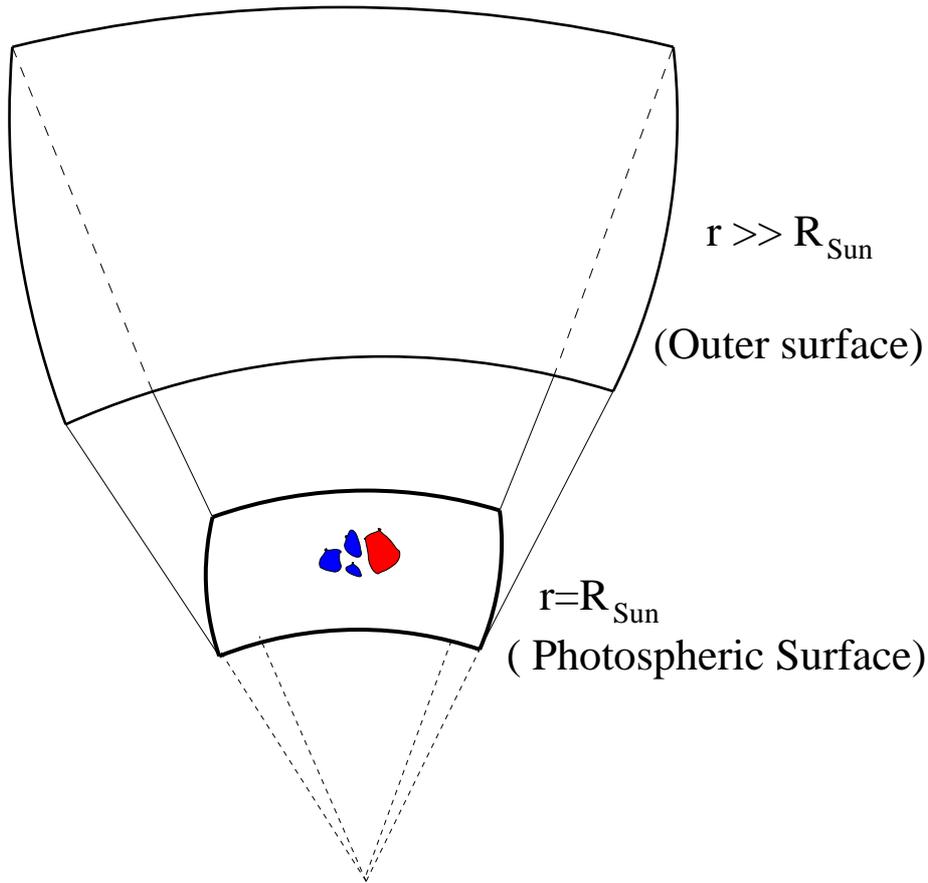}
              }
   \caption{
   Schematic illustration of the volume in which the photospheric-to-coronal
   portions of a bipolar active region are imbedded.  It is assumed that at the
   outer surface, the magnetic field is negligibly small, and that the 
   side-wall 
   boundaries are sufficiently far away from the active region that they
   do not contribute to the Gauss' theorem surface integral.  Note that
   at the photosphere, the outward surface normal vector $\vecnhat$ 
   points in the $-\vecrhat$ direction.  The red and blue colors
   represent the upward and downward vertical fluxes in the active region.
   }
   \label{figure:spherevolume}
   \end{figure}

The Lorentz force per unit volume can be written as
\be
\vecf_L = \grad \cdot \vecT = 
{\partial T_{ij} \over \partial x_j}\ ,
\label{equation:maxuse}
\ee
where the Maxwell stress tensor [$T_{ij}$] is given by
\be
T_{ij} = {1 \over 8 \pi} ( 2 B_i B_j - B^2 \delta_{ij} ) \ ,
\label{equation:maxdef}
\ee
and $B_i$ and $B_j$ each range over the three components of the magnetic 
field [$\vecB$], and $\delta_{ij}$ is the Kronecker delta function.  Here,
the divergence is expressed in Cartesian coordinates.
To evaluate the total Lorentz force [$\vecF_L$]
acting on the atmospheric volume
that surrounds a flaring active region,
we integrate this force density over the volume,
with the photospheric surface taken as 
the lower boundary of that volume, and with the upper boundary taken at some
great height above the active region.  The volume integral of the divergence
in Equation (\ref{equation:maxuse}) can be evaluated by using the 
divergence (Gauss') theorem:
\be
\vecF_L \equiv \int_V \mathrm{d}^3x {\partial T_{ij} \over \partial x_j} =
\oiint_{A_\mathrm{tot}} dA\, T_{ij} n_j\ ,
\label{equation:gaussthmtens}
\ee
where $n_j$ represents the components of the outward unit vector
[$\vecnhat$] that is normal to the bounding surface of the atmospheric volume,
and where $A_\mathrm{tot}$ represents the area of the entire bounding surface.
Substituting the expression (\ref{equation:maxdef})
for the Maxwell stress tensor, results in this equation:
\be
\vecF_L = {1 \over 8 \pi} 
\oiint_{A_\mathrm{tot}} \mathrm{d}A\, \left[ 2 \vecB (\vecB \cdot \vecnhat) - 
\vecnhat B^2 \right] \ .
\ee
A sketch of the atmospheric volume surrounding the active region is shown
in Figure \ref{figure:spherevolume}.

If we assume that the upper surface of the volume is sufficiently far
above the active region that the magnetic field integrated over that
surface is negligible, and that the side walls are also sufficiently distant
that there is negligible magnetic field contribution from those
integrals as well, then the only surface that will contribute will be
the photosphere near the active region, 
where the magnetic fields are strong.  In that
case, $\vecnhat = -\vecrhat$, and $\vecB \cdot \vecnhat = - B_r$, where $B_r$ is
the radial field component.  The surface integral then
results in the following two equations for 
the upward ($i.e.$ radial) and horizontal
components of the Lorentz force [$F_r$] and [$\vecF_h$]:
\be
F_r = {1 \over 8 \pi} \int_{A_\mathrm{ph}} \mathrm{d}A\, ( B_h^2 - B_r^2 )\ ,
\label{equation:forceup}
\ee
and
\be
\vecF_h = -{1 \over 4 \pi} \int_{A_\mathrm{ph}} \mathrm{d}A\, B_r \vecB_h\ .
\label{equation:forcesideways}
\ee
Here, $\vecB_h$
represents the components of $\vecB$
in the directions parallel to the photosphere, which we will henceforth
refer to as the ``horizontal'' directions.  The quantity 
$B_h^2 = \vecB_h \cdot \vecB_h$,
and $A_\mathrm{ph}$ is the area of the photospheric
domain containing the active region.
If the active region is sufficiently small in spatial extent and magnetically
isolated from other strong magnetic fields, 
one can approximate this surface integral as an integral
over $x$ and $y$ in Cartesian coordinates,
with the upward direction represented as $z$ instead of $r$.  

The restrictions given above regarding the side-wall contributions to
the Gauss's law integrals can be relaxed if the volume of the domain
is extended to a global volume:
an integral over the entire outer atmosphere of the Sun.  In this case, there
are no side-wall boundaries to worry about, and $A_\mathrm{ph}$
coincides with the entire photospheric surface of the Sun.  The outer
spherical surface 
boundary is assumed to be sufficiently far from the Sun
that it makes no significant 
contribution to the Gauss's law surface integral.  Thus,
integrals over the entire solar surface of Equations (\ref{equation:forceup}) 
and (\ref{equation:forcesideways}) should represent 
the total Lorentz force acting on the Sun's outer atmosphere.
If one sets the total Lorentz force to zero, the above surface integrals
yield well-known constraint equations on force-free fields 
(\opencite{Low1985}), 
with the Cartesian version of the equations being used to test the force-free
condition of photospheric and chromospheric vector magnetograms
(\opencite{Metcalf1995}).  If the magnetic-field distribution is not
force-free, but the atmosphere is observed to be
static, then presumably the Lorentz forces
are balanced by other forces such as gas-pressure gradients and gravity.

\inlinecite{Wang2010} have found from an analysis of eleven large (X-class)
flaring active
regions that the vector magnetic field is always
observed to change after a flare
in the sense that the magnetic
field becomes ``more horizontal'' than
before the flare.  The change is observed 
to occur on a timescale of a few
minutes, 
and in some cases on a time scale as fast as the sample spacing 
(one minute) permits.
What are the implications of this observational result
on the Lorentz force acting on the solar atmosphere?

To address this question, we first take the temporal derivative of Equations
(\ref{equation:forceup}) and (\ref{equation:forcesideways}) to find
\be
{\partial F_r \over \partial t} = {1 \over 8 \pi} \int_{A_\mathrm{ph}}
\mathrm{d}A\, {\partial \over \partial t} (B_h^2 - B_r^2)\ ,
\label{equation:dforceupdt}
\ee
and
\be
{\partial \vecF_h \over \partial t} = -{1 \over 4 \pi} 
\int_{A_\mathrm{ph}} 
\mathrm{d}A\, {\partial \over \partial t} ( B_r \vecB_h )\ .
\label{equation:dforcesidewaysdt}
\ee
Next, we assume the fields are observed to change over a time duration
$\delta t$, and integrate the temporal derivatives of the Lorentz-force
contributions to find the changes to the Lorentz-force components
$\delta F_r$ and $\delta \vecF_h$:
\be
\delta F_r = {1 \over 8 \pi} \int_{A_\mathrm{ph}} \mathrm{d}A\,
( \delta B_h^2 - \delta B_r^2 )\ ,
\label{equation:deltafr}
\ee
and
\be
\delta \vecF_h = -{1 \over 4 \pi} \int_{A_\mathrm{ph}} \mathrm{d}A\,
\delta ( B_r \vecB_h )\ ,
\label{equation:deltafh}
\ee
where at a fixed location in the photosphere
\bea
\delta B_h^2 &\equiv& \int_0^{\delta t} \mathrm{d}t\,
{\partial \over \partial t} B_h^2
= B_h^2(\delta t) -B_h^2(0)\ ,\\
\delta B_r^2 &\equiv& \int_0^{\delta t} \mathrm{d}t\,
{\partial \over \partial t} B_r^2
= B_r^2(\delta t) -B_r^2(0)\ ,\\
\delta ( B_r \vecB_h ) &\equiv& 
\int_0^{\delta t} \mathrm{d}t\,{\partial \over \partial t} ( B_r \vecB_h )
= B_r(\delta t)\vecB_h(\delta t) - B_r(0) \vecB_h(0)\ .
\eea
These quantities are simply the observed changes in
the magnetic variables that occur over the course of a flare.  
Note that if
the flaring active region is near disk center, so that the observed transverse
magnetic field is a good approximation to the horizontal field, then
the expression for $\delta F_r$ can be evaluated without having to perform
the 180$^{\circ}$ disambiguation
of the vector magnetogram data -- only the amplitude of the horizontal
field enters into the expression.

If the change in the Lorentz force is significant only within small areas
of the photosphere, 
then the only contribution to the
global-area integral 
will be from
the smaller domains where the changes are significant,
potentially simplifying the evaluation of Equations (\ref{equation:deltafr}) and
(\ref{equation:deltafh}).

We now argue that in the outer atmosphere of flaring active regions, 
the impulse from the changed Lorentz force dominates all other forces.
First, from energetic considerations, the magnetic field is believed to be
the source of energy for eruptive flares and coronal mass ejections:
\inlinecite{Forbes2000} and \inlinecite{Hudson2007} have 
argued that no other known source of energy can provide the
observed kinetic energy of outward motion observed in coronal mass ejections,
and there simply is no other viable source for the thermal and radiated
energy known to be released in solar flares.  
Second, apart from the Lorentz force, the only other significant forces
known to be operating on the solar atmosphere are gas-pressure gradients
and gravity.  To evaluate the change in the gas-pressure gradient forces in
the outer atmosphere, one
can perform a Gaussian volume integral over the outer solar atmosphere
of the vertical component of
the gas-pressure gradient force.   The net change in the vertical force is just
the difference between the gas pressure change at the top of the Gaussian volume
from that at the bottom.  If the plasma $\beta$ in the solar atmosphere
is low, as is generally the case in active regions, it seems unlikely
that this will be as significant as the change of the Lorentz force.
Nevertheless, in Section \ref{section:otherforce}, we will consider 
perturbations to the gas pressure at the photosphere and discuss
their effectiveness.
In the case of the gravitational force, unless the plasma has moved a huge
distance ($\approx R_{\odot}$) away from the Sun on the 
time-scale of the observed field change,
the gravitational force acting on the given mass of the plasma within the
Gaussian volume must be approximately the same, and hence the change in
the gravitational force should be small.

The results of \inlinecite{Wang2010}, in which the
field is observed to become more horizontal after the occurrence of eruptive
flares, is thus
consistent with an upward impulse acting on the outer atmosphere, 
so we identify this impulse as the photospheric magnetic-field signature of 
the force driving
a magnetic eruption.  To estimate the magnitude of the impulse, we
make the simple assumption that the change in the Lorentz force in Equations
(\ref{equation:deltafr}) and (\ref{equation:deltafh}) occurs linearly
with time
from $t=0$ to $t=\delta t$.  
We denote the mass of
the plasma that is eventually ejected as $M_\mathrm{ejecta}$, and we assume
the fluid velocity averaged over this plasma is zero prior to the
eruption.  The Lorentz impulse will then be related to the ejecta's
momentum by
\be
{1 \over 2}\, \delta F_r\, \delta t = M_\mathrm{ejecta} v_r
\label{equation:impulser}
\ee
and
\be
{1 \over 2}\, \delta \vecF_h\, \delta t = M_\mathrm{ejecta} \vecv_h\ ,
\label{equation:impulseh}
\ee
where $v_r$ is the upward (radial) component of the velocity of the ejecta
after the impulse, and $\vecv_h$ is the resulting horizontal component of 
the ejecta velocity.  Note that if $v_r$ is less than the escape velocity
[$v_\mathrm{e} \equiv (2 G M_\odot / R_\odot)^{1/2}$],
the ejecta will ultimately be stopped by gravity and will not
result in an eruption.  If $v_r$ exceeds $v_e$, we
assume that the ejecta can become a coronal mass ejection (CME).
This means that for
a given observation of magnetic-field changes in a flaring active region, 
there is an upper limit to the mass of any resulting CME given by
\be
M_\mathrm{CME} < {1 \over 2}\, {\delta F_r\, \delta t \over
v_\mathrm{e}}\ .
\label{equation:massul}
\ee
For the magnetic-field changes in the 2 November 2003 flare
studied by \inlinecite{Hudson2008}, they
estimate a change in the Lorentz-force surface density of 
$2500$ ${\rm dyne\ cm}^{-2}$, and with the estimated area over which the change
occurs, a total force of $\approx 10^{22}\ {\rm dyne}$.  This
is probably an underestimate for the total upward force, 
since this case was taken from the study by \inlinecite{Sudol2005}, in
which only the line-of-sight contributions were measured.
A more extensive analysis of a larger dataset of flares
(\opencite{Petrie2010}) has since shown several other cases 
of comparable or even larger Lorentz forces for some X-class flares.
Assuming
a time scale of ten minutes for the photospheric magnetic fields to change 
(from the temporal evolution results of \opencite{Sudol2005} and
\opencite{Petrie2010}) then
results in an upper limit on the mass of any CME coming from this flaring active
region of $4.9 \times 10^{16}\ {\rm g}$.

Having an expression for the Lorentz-driven impulse in the horizontal directions
(Equation (\ref{equation:impulseh})) could
be useful in determining the initial deflection of the ejecta away from a radial
trajectory.  The initial trajectory direction can be determined by examining
the ratio of the horizontal components of $\delta \vecF_h$ to $\delta F_r$.

What effect do these Lorentz force changes, and the impulses driven by them,
have on the response of the solar interior plasma at and below the photosphere?
To estimate the change in the Lorentz force acting on the solar interior
(the interior is defined here to be
the plasma that extends from the photosphere downward),
one can perform almost exactly the
same Gaussian volume exercise as above, but using a sub-surface volume instead
of an outer atmosphere volume.  By performing the global integral of the
Lorentz-force density over the entire volume below the photosphere, 
one can see that both the absolute Lorentz forces 
(Equations (\ref{equation:forceup}) and 
(\ref{equation:forcesideways})) and the changes in
the Lorentz forces (equations (\ref{equation:deltafr}) and 
(\ref{equation:deltafh})) involve exactly the same photospheric
surface terms as for the outer solar atmosphere, except that the outward surface
normal $\vecnhat$ is in the $+\vecrhat$ direction instead of in the
$-\vecrhat$ direction.  Thus the three components of the Lorentz
force, and the flare-induced changes in the Lorentz force, have exactly the
same magnitude, but opposite sign from the Lorentz forces acting on the solar
atmosphere -- the Lorentz force changes acting 
on the interior and the solar atmosphere
are exactly balanced:
\be
\delta F_{r, \mathrm{interior}} = {1 \over 8 \pi} 
\int_{A_\mathrm{ph}} \mathrm{d}A\,
( \delta B_r^2 - \delta B_h^2 )\ ,
\label{equation:deltafrdown}
\ee
and
\be
\delta \vecF_{h, \mathrm{interior}} = {1 \over 4 \pi} 
\int_{A_\mathrm{ph}} \mathrm{d}A\,
\delta ( B_r \vecB_h )\ .
\label{equation:deltafhdown}
\ee
The radial component of the
force change acting on the solar interior was
identified as a magnetic jerk by \inlinecite{Hudson2008}. 

To relate these results 
(Equations \ref{equation:deltafrdown}\,--\,\ref{equation:deltafhdown}) 
to those of
\inlinecite{Hudson2008}, we note that if we let $z$ be in the upward direction,
and use $x$ and $y$ to denote the horizontal directions, and further make
the first order approximation that
$\delta B_h^2 \approx 2 B_x \delta B_x + 2 B_y \delta B_y$, and that
$\delta B_z^2 \approx 2 B_z \delta B_z$, where 
$\delta B_x,\ \delta B_y,\ \delta B_z$
are the observed changes in $B_x$, $B_y$, and $B_z$,
then Equation (\ref{equation:deltafrdown}) yields the un-numbered 
expression given by \inlinecite{Hudson2008}, assumed to be
integrated over the vector-magnetogram area:
\be
\delta F_{z, \mathrm{interior}} \approx {1 \over 4 \pi} \int\,\mathrm{d}A\, 
(- B_x\ \delta B_x - B_y \delta B_y + B_z \delta B_z )\ .
\label{equation:hfw}
\ee
If Equation (\ref{equation:hfw}) is evaluated over the
flaring active region,
such that surface terms on the vertical side walls make no
significant contributions to the Gaussian integral, and the amplitude
of the field-component changes is small compared to their initial values,
then this expression should be robust and accurate.
For future investigations of vector-magnetogram data, we believe that 
Equations (\ref{equation:deltafrdown})\,--\,(\ref{equation:deltafhdown}) will 
generally be more useful than Equation (\ref{equation:hfw})
since they do not assume the first-order appoximation, and the horizontal
components of the force are included.

Since we assert that the Lorentz-force change
is the dominant force acting in the outer solar atmosphere, and that this
force drives an eruptive impulse, it follows
from conservation of momentum that an equal and opposite impulse must be
applied on the plasma in the solar interior, 
and that at least initially, the force driving this impulse is 
the Lorentz force identified in Equations 
(\ref{equation:deltafrdown}) and (\ref{equation:deltafhdown}).
However, we expect 
that once the impulse has penetrated more than a few pressure scale heights
into the solar interior, that the disturbance will propagate mainly as
a gasdynamic-pressure disturbance (acoustic wave), since 
the plasma $\beta$ is thought to
increase very rapidly with depth below the photosphere.
For a more general discussion about momentum balance issues in solar flares,
see \inlinecite{Hudson2011}, included in this topical issue.

Putting all of this together, we suggest that there is an observationally
testable relationship between the measured Lorentz-force change and the outward
momentum of the erupting ejecta that occurs over the course of an eruptive 
flare, and that the Lorentz force responsible for the eruption should also drive
a downward-moving impulse into the solar interior. 
The downward-moving impulse could potentially be the source
of observed ``sunquake'' acoustic emission detected with helioseismic
techniques 
(\opencite{Kosovichev1998}; \opencite{Moradi2007}; \opencite{Kosovichev2011})
for some solar flares.  Thus we suggest the possibility
of using helioseismology to study eruptive solar flares, if the detailed
wave mechanics of the impulse moving downward into the interior can be
better characterized and better understood.

\section{Other Disturbances in the Force}
\label{section:otherforce}

We argue above that assuming that the plasma $\beta$ in the flaring
active region is small implies that changes to gas-pressure gradients during
a flare are most likely unimportant
compared to changes in the Lorentz force.  Nevertheless, 
the flare-induced gas-pressure change from energy 
deposited in the flare atmosphere
has been considered in the past to be
a viable candidate for the agent that excites flare-associated
helioseismic disturbances (\opencite{Kosovichev1995}).  
Another suggested mechanism is heating
near the solar photosphere driven by radiative backwarming of strong
flaring emission occurring higher up in 
the solar atmosphere (\opencite{Donea2006}; \opencite{Lindsey2008}; 
\opencite {Moradi2007}).  We
consider each of these possibilities in the following sections.

\subsection{Pressure Changes Driven by Flare Gasdynamic Processes}
\label{section:gasbag}

During the impulsive phase of flares,
emission in the hard X-ray and $\gamma$-ray
energy range is typically emitted from 
small, rapidly moving kernels in the chromosphere of the flaring active region
($e.g.$ \opencite{Fletcher2007}).
This emission is generally assumed to be the signature of energy release
in the form of a large flux of energetic electrons.  Energetic electrons
in the 10\,--\,100 KeV range that impinge on the solar atmosphere will 
emit nonthermal bremsstrahlung radiation from Coulomb collisions with the
ambient ions in the atmosphere, and will also rapidly lose energy
via Coulomb collisions with ambient electrons, resulting in strong
atmospheric heating (\opencite{Brown1971}).  This
results, in turn, in a large gas pressure increase 
in the upper chromosphere, due to rapid chromospheric evaporation.
\inlinecite{Kosovichev1995} proposed that this 
large pressure increase is the agent responsible for
flare-driven helioseismic waves into the solar interior that have been observed.

Can this pressure increase in the flare chromosphere result in a
sufficiently great pressure change at the photosphere to be significant compared
to the observed changes in the Lorentz force?
To investigate this question, 
we show that any gasdynamic disturbance that reaches the photosphere will
propagate first as a shock-like disturbance (a ``chromospheric condensation'')
followed by the propagation of a weaker disturbance that can be treated in
the acoustic limit ($v / c_s \ll 1$), and we then describe the work
necessary to determine whether this acoustic disturbance 
has a perturbed pressure that can be
comparable in strength to the Lorentz-force disturbance we considered in 
Section \ref{section:maxwell}.  
The first task is to estimate the pressure increase
in the flare chromosphere that drives the chromospheric condensation.

The pressure increase in the flare chromosphere occurs when plasma that was
originally at chromospheric densities is heated to 
coronal temperatures.  The size of the pressure increase depends on the
details of how the heating is applied to the pre-flare atmosphere.  If
the flux of non-thermal electrons is increased very suddenly, and with
a sufficiently great amplitude to exceed the maximum
ability of the upper chromosphere to radiate away the non-thermal electron
energy flux, this results in ``explosive evaporation'' 
(\opencite{Fisher1985a}, \opencite{Fisher1985b}).  
In this case, the location of the flare
transition region moves very quickly to a significantly greater depth in the
atmosphere.  The column depth of this location can be determined by applying
the suggestion of \inlinecite{Lin1976}, equating the non-thermal electron
heating rate with the maximum radiative cooling rate, assuming a transition
region temperature.  The validity of this approach was
subsequently verified in the numerical simulations
of \inlinecite{Fisher1985a}.  The plasma between the original and flare
transition region column depths then explodes, driving violent mass motion
both upwards and downwards.  

\begin{figure}    
   \centerline{\includegraphics[width=1.0\textwidth,clip=]
   {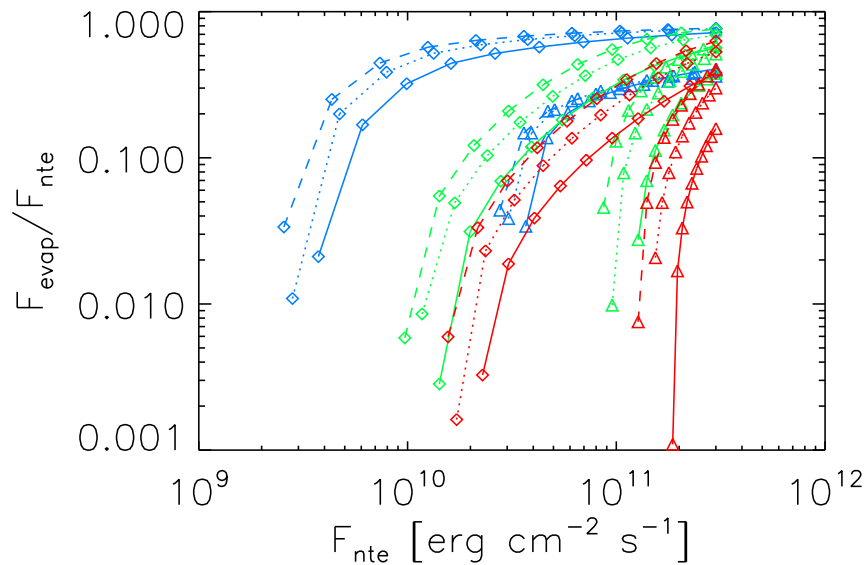}
              }
   \caption{
   Fraction of the total non-thermal electron energy flux [$F_\mathrm{nte}$]
   that goes into driving
   explosive chromospheric evaporation as a function of $F_\mathrm{nte}$.
   Diamonds indicate a pre-flare coronal pressure of
   0.3 ${\rm dyne}\ {\rm cm}^{-2}$, 
   while triangles indicate a higher
   pre-flare coronal pressure of 3.0 ${\rm dyne}\ {\rm cm}^{-2}$.
   Blue symbols indicate a 10 KeV low-energy cutoff, green indicates
   a 20 KeV low-energy cutoff, and red indicates a
   25 KeV low-energy cutoff.  
   Solid curves indicate an electron spectral index
   $\delta = 4$, dotted curves $\delta = 5$, and dashed
   curves $\delta = 6$.
   }
   \label{figure:fratioplot}
   \end{figure}

\begin{figure}    
   \centerline{\includegraphics[width=1.0\textwidth,clip=]
   {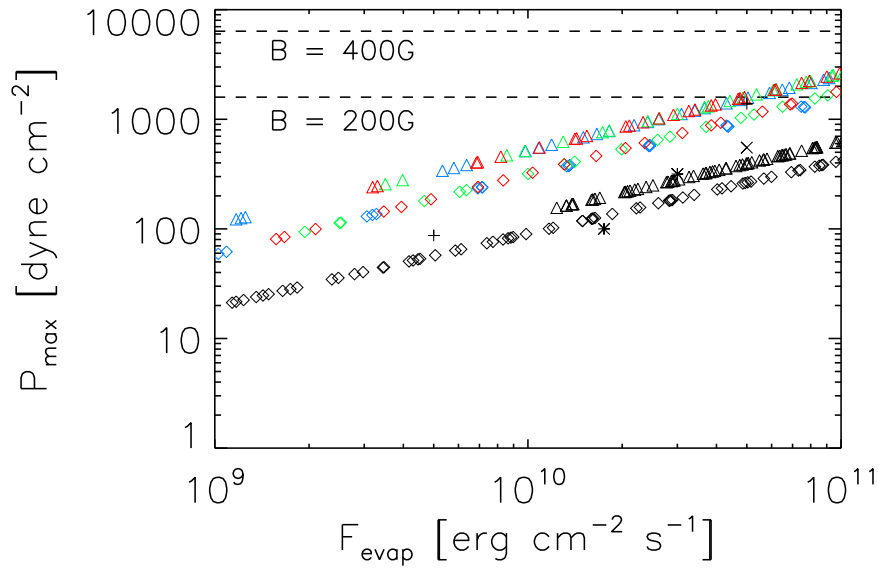}
              }
   \caption{
  Maximum gas pressure driven by impulsive-phase flare heating from non-thermal
  electrons as a function of the energy flux driving chromospheric evaporation.
  The colored diamonds show the maximum pressure computed using the analytical
  explosive evaporation model of Fisher (1987), for a pre-flare coronal
  pressure of 0.3 ${\rm dyne}\ {\rm cm}^{-2}$,
  while the triangles show the the maximum pressure for a pre-flare coronal
  pressure of 3.0 ${\rm dyne}\ {\rm cm}^{-2}$.
  Blue symbols indicate a 10 KeV low-energy cutoff, green indicates
  a 20 KeV low-energy cutoff, and red indicates a
  25 KeV low-energy cutoff.
  The black symbols indicate the maximum
  pressure using the alternative
  model that the non-thermal energy drives evaporation indirectly via thermal
  conduction from Equation (32) of Fisher (1989).
  The black diamonds and triangles indicate the same preflare coronal
  pressures as above.
  Note both approximations show that
  $P_\mathrm{max} \approx F_\mathrm{evap}^{2/3} \rho_\mathrm{co}^{1/3}$,
  where $\rho_\mathrm{co}$ is
  the preflare coronal mass density, but with different proportionality
  constants.
  Also plotted are the maximum pressures from several radiation-hydrodynamic
  flare simulations:  The two
  highest flux cases in Fisher, Canfield, and McClymont
  (1985a,b,c) (*),
  two cases from Abbett and Hawley (1999) (+), and a case
  from Allred {\it et al.} (2005) ($\times$).  The two dashed horizontal
  lines denote the magnetic pressure for field strengths
  of 200\,G and 400\,G, respectively.
  }

   \label{figure:pmaxplot}
   \end{figure}

\inlinecite{Fisher1987} developed an analytical
model for the explosive evaporation process, including estimates for the 
maximum pressure achieved during explosive evaporation, in terms of the 
portion of the total non-thermal electron energy flux, $F_\mathrm{evap}$, 
that is deposited between the original and flare
transition-region column depths.  In Figure \ref{figure:fratioplot}, we 
explore explosive evaporation by first showing
the computed
ratio of $F_\mathrm{evap}$ to the total flux of non-thermal electrons
$F_\mathrm{nte}$
for many possible
cases of explosive evaporation, assuming a range of preflare atmospheric
coronal pressure, values of the assumed low-energy cutoff [$E_c$], the electron
spectral index [$\delta$]
(see Section III of \inlinecite{Fisher1985a} for definitions of $E_c$ and $\delta$), 
and the total flux of non-thermal electrons
$F_\mathrm{nte}$.  Preflare
coronal pressures include a low value of
$0.3\ \mathrm{dyne\ cm}^{-2}$ (diamonds), corresponding to a tenuous 
pre-flare coronal density,
and a higher value of
$3.0\ \mathrm{dyne\ cm}^{-2}$ (triangles), corresponding to a denser pre-flare
corona.  
Values of $E_c$ include $10\ \mathrm{KeV}$ 
(blue), $20\ \mathrm{KeV}$ (green), and $25\ \mathrm{KeV}$ (red).  Electron
spectral index [$\delta$] values include 4 (solid curves), 5 (dotted curves), 
and 6 (dashed curves).
Rather than assuming a sharp low-energy cutoff to the spectrum, which produces
an unphysical cusp in the non-thermal-electron heating rate as a function
of column depth, we adopt the modified form of the heating rate suggested in
Figure 1 and Equations (9)\,--\,(11) of
\inlinecite{Fisher1985a}, in which the heating rate varies smoothly with column
depth.  This implies the low-energy cutoff [$E_c$] corresponds more to a
spectral rollover than a true cutoff; the detailed
electron spectra corresponding to this
particular rollover behavior is given in Equations (46)\,--\,(50) 
of \inlinecite{Tamres1986}.  Once the values for $F_\mathrm{evap}$ have been
obtained, one can find the average per-particle heating rate in the
explosively evaporating region and use Equations (38)\,--\,(39) from 
\inlinecite{Fisher1987} to compute the maximum pressure due to explosive
evaporation.  The resulting values are shown as the colored triangles and
diamonds in Figure \ref{figure:pmaxplot} as functions of the flux of energy
driving explosive evaporation.

We also compute the maximum pressure using an entirely different assumption
for how chromospheric evaporation occurs.  If the energy flux of 
non-thermal electrons
increases more slowly than the time scale for which explosive evaporation
occurs, or if the non-thermal electron energy is simply dumped
into the corona and transition
region through bulk heating, then chromospheric evaporation will still occur,
and the pressure will still increase, but not as violently as assumed
in the model of \inlinecite{Fisher1987}.  
In the Appendix of \inlinecite{Fisher1989a}, it was shown
by considering the dynamics of chromospheric evaporation
driven via thermal conduction that 
the maximum pressure is given approximately by
Equation (32) of \inlinecite{Fisher1989a}.  We use this expression to 
compute a second estimate for the maximum pressure driven by chromospheric
evaporation, using the energy flux deposited above the flare transition
region from the assumed atmospheric and electron spectral characteristics
described earlier.  The results are shown as the black diamonds and triangles
in Figure \ref{figure:pmaxplot}.

Note that both sets of estimates for the maximum pressure result in the
scaling $P_\mathrm{max} \sim F_\mathrm{evap}^{2/3} \rho_\mathrm{co}^{1/3}$,
where $\rho_\mathrm{co}$ is the preflare coronal mass density.
This result is consistent with what one might find from a simple
dimensional analysis; the only difference between the two estimates is simply a
different constant of proportionality that results from the detailed
assumptions in the two different evaporation models.

In addition to these estimates described above, 
we also plot in Figure \ref{figure:pmaxplot}
the maximum pressure as a function of the estimated energy flux driving
evaporation achieved in the two largest-flux simulations
of \inlinecite{Fisher1985a}, two cases from \inlinecite{Abbett1999}, and
the highest-flux case from \inlinecite{Allred2005}.  Note that in all cases,
the results from the gasdynamic simulations are either close to the
approximated maximum pressures from the estimates derived above, or else are
bracketed by our estimates.  This is even true when the assumptions used
in deriving the approximate results are not strictly adhered to in the
simulations.  We therefore can feel some confidence that our simpler estimates
will probably bracket most cases.  In particular, the explosive evaporation
estimates (colored triangles and diamonds) seem to
provide good upper limits to
the maximum pressure due to chromospheric evaporation found from any of
the simulations.  By examining Figures
\ref{figure:fratioplot} and
\ref{figure:pmaxplot}, we conclude that for large flare
non-thermal electron energy fluxes
$\approx 10^{11} {\rm\ erg\ s}^{-1} {\rm \ cm}^{-2}$,
the maximum pressure increase in the chromosphere 
is $\approx 2000\ {\rm\ dyne\ cm}^{-2}$, and in most cases, considerably less
than this.  In summary, to achieve gas-pressure increases this high requires the
highest non-thermal energy fluxes and a very rapid onset of these high
energy fluxes.

The pressure increase drives not only the rapid upward motion of the evaporating
plasma into the corona, but also slower, denser flows
of plasma downward into the chromosphere ($e.g.$ \opencite{Fisher1985c})
described as ``chromospheric condensations''.
These dense, downward-moving plugs of plasma form
behind a downward-moving shock-like disturbance driven 
by the pressure increase from chromospheric evaporation in flares.  
Simple, analytic
models of the dynamic evolution of chromospheric condensations
were developed by \inlinecite{Fisher1989a}. The models were found to do a good
job of describing the results of more detailed numerical gasdynamic simulations.
One interesting property of the models is that, during the time period
that the downflow evolution is well-described in terms of chromospheric
condensations, the dynamical evolution is insensitive to the details
of cooling behind the downward moving front of the chromospheric condensation.
Further, examination of several simulation results indicate that the gas
pressure in the chromospheric condensation just behind the front of the
condensation is relatively constant in time, as the condensation
propagates deeper into the atmosphere.  This result was used in the
analytical models of the condensation dynamics (\opencite{Fisher1989a}).
Radiative cooling immediately behind the downward moving
shock at the head of the chromospheric
condensation leads to densities in the condensation that are much greater 
than the density ahead of it.  \inlinecite{Fisher1989a} showed by applying 
mass- and momentum-conservation jump conditions, plus differing assumptions
about how the plasma is cooled, that the velocity evolution is
very insensitive to the details of how the plasma is cooled, 
provided that the resulting
density jump is large (see the comparisons in Figure 1 of that article, 
for example).  
These models predict 
the maximum column depth that the chromospheric condensation
can penetrate into the solar atmosphere as a shock-like disturbance, 
in terms of the flare-induced
pressure [$P_\mathrm{max}$] driven by electron-beam heating 
of the solar atmosphere.  The
maximum column depth of propagation [$N_\mathrm{max}$] is 
given approximately by
\be
N_\mathrm{max} = { P_\mathrm{max} \over \bar m g}\ ,
\label{equation:nmax}
\ee
where $\bar m \approx 1.4 m_p$ is the mean mass per proton in the solar
atmosphere, and $g = 2.74 \times 10^4 {\rm\ cm\ s}^{-2}$ is 
the value of surface gravity.
The values of $P_\mathrm{max}$ mentioned above
result in values of $N_\mathrm{max}$ that
are no larger than $\approx 3 \times 10^{22}\ {\rm \ cm}^{-2}$.  
The column depth of the
solar photosphere, on the other hand is $\approx 10^{24} {\rm \ cm}^{-2}$.
Thus, using the chromospheric-condensation model, flare-driven pressure
disturbances can propagate to at most 3\% of the column depth of the
photosphere as chromospheric condensations;
but this does not mean that the downflows cease at this
depth: it means only that the equation of motion for 
chromospheric condensation (Equation (10) from \inlinecite{Fisher1989a}) 
no longer applies when the driving pressure approaches
the ambient pressure ahead of the condensation.  At the depths where 
this occurs, downflow velocities become significantly less than the sound speed,
and are therefore better treated in the acoustic limit.

Because the chromospheric-condensation model's assumptions begin to break
down at the last stages of its evolution, we then consider the subsequent
downward
propagation of flare-driven pressure disturbances between column depths
of $\approx 3 \times 10^{22}\ {\rm \ cm}^{-2}$ and 
$\approx 10^{24} {\rm \ cm}^{-2}$
using an entirely
different approach:  We assume that the disturbance can be represented
by an acoustic wave, driven by a simple downward pulse corresponding to the last
stages of the chromospheric condensation's evolution.
For simplicity, we assume simple, adiabatic wave evolution in
an isothermal, gravitationally stratified approximation of the lower
chromosphere, assuming an ideal gas equation of state,
without dissipation.  As described in more detail below,
there are reasons to question these assumptions, but this solution allows us
to demonstrate some general properties of the resulting wave evolution.

Assuming that the pre-flare chromosphere can be represented by an isothermal,
gravitationally stratified
atmosphere at temperature $T_\mathrm{ch}$ with pressure scale height
$\Lambda_P = c_s^2 / (\gamma g )$, where $c_s$ is the adiabatic sound speed,
and $\gamma$ is the ratio of specific heats,
the equation for the perturbed vertical velocity is found to be
\be
{\partial^2 v \over \partial t^2} - c_s^2 ({1 \over \Lambda_P} {\partial v \over
\partial s} + { \partial^2 v \over \partial s^2}) = 0\ .
\label{equation:acoustic}
\ee
Here, $s$ measures vertical distance in the downward direction, measured from
the final position of the chromospheric condensation.
We assume that at the depth where the chromospheric-condensation solution
breaks down, the result of its final propagation is a
downward displacement [$\Delta s$] occurring over a short time.  
We then want to
follow this displacement, using the above acoustic-wave equation, as it
propagates downward.  At $s=0$, we therefore assume the temporal evolution of
the velocity [$v$] is given by 
\be
v(s=0,t) = \Delta s \delta(t)\ ,
\label{equation:initialwave}
\ee
where $\delta(t)$ is the Dirac $\delta$-function.  By performing 
a Laplace transform
of Equation (\ref{equation:acoustic}) with this assumed time behavior at
$s=0$, we find
\bea
&&v(s,t) = \Delta s\ {\rm exp}\left(-{s \over 2 \Lambda_P}\right)\nonumber \\ 
&&\times \left[ \delta(t-{s \over c_s}) - 
{s \over 2 \Lambda_P \sqrt{t^2-{s^2 \over c_s^2}}} 
J_1 \left( \omega_a \sqrt{t^2 - {s^2\over c_s^2}} \ \right) 
H(t-{s \over c_s}) \right]\ ,
\label{equation:acousticsolution}
\eea
where $J_1$ is a Bessel function, $H$ is the Heaviside function, and 
$\omega_a$
is the acoustic cutoff frequency [$\omega_a = c_s / (2 \Lambda_P)$].

Note that the solution corresponds to the downward propagation of the pulse,
along with a trailing wake that oscillates at a frequency that
asymptotically approaches the acoustic-cutoff frequency.  While it is clear that
the velocity amplitude decreases rapidly (there is an overall envelope function
going as $\exp(-s / (2 \Lambda_P))$ as the pulse propagates deeper),
one can show that the perturbed pressure amplitude associated with the
pulse actually increases as $\exp(s / (2 \Lambda_P))$
as the pulse propagates deeper.
It is therefore possible that the pressure amplitude of the acoustic wave could
be even larger than the initial pressure [$P_\mathrm{max}$] of the flare
chromosphere, derived above.

On the other hand, there are a number of dissipation
mechanisms that our wave solution does not include, 
which could dramatically reduce the perturbed
pressure of the resulting acoustic wave.
To the extent that energy balance between radiative cooling and flare heating
by the most energetic electrons is important at these depths,
\inlinecite{Fisher1985c} showed in Section IV of that article 
that high-frequency acoustic waves were
very strongly damped by radiative cooling; low-frequency acoustic waves were
damped more weakly, but still attenuated on length scales of $\approx 200$ km.

To summarize, we have shown how one can estimate the peak gasdynamic 
pressure driven by chromospheric evaporation, and that the largest 
possible values of this pressure require both high energy fluxes and 
rapid onset of those fluxes to produce explosive evaporation.  
We have also shown that the ensuing shock- like disturbance 
(a chromospheric condensation) can only propagate down to 
roughly 3\% of the column depth of the photosphere, but that the 
dying chromospheric condensation can continue propagating downward as 
an acoustic wave to photospheric depths. We found a wave solution 
that is dispersive, consisting of both a pulse and a trailing wake, 
and used a simplified example to show that the perturbed pressure of 
the acoustic wave can increase as the wave propagates down towards 
the photosphere.
We then
discuss a number of wave-dissipation mechanisms that
may efficiently extract energy from the wave.  The extent to which
the gas-dynamic-excited acoustic wave at the photosphere
is important relative to the Lorentz-force perturbation 
(Section \ref{section:maxwell})
will depend critically on a detailed evaluation of these 
wave-dissipation effects on
the acoustic solution, and it is beyond the scope of what we can present here.

\subsection{Pressure Changes Driven by Radiative Backwarming}
\label{section:backwarming}

\begin{figure}    
   \centerline{\includegraphics[width=1.0\textwidth,clip=]{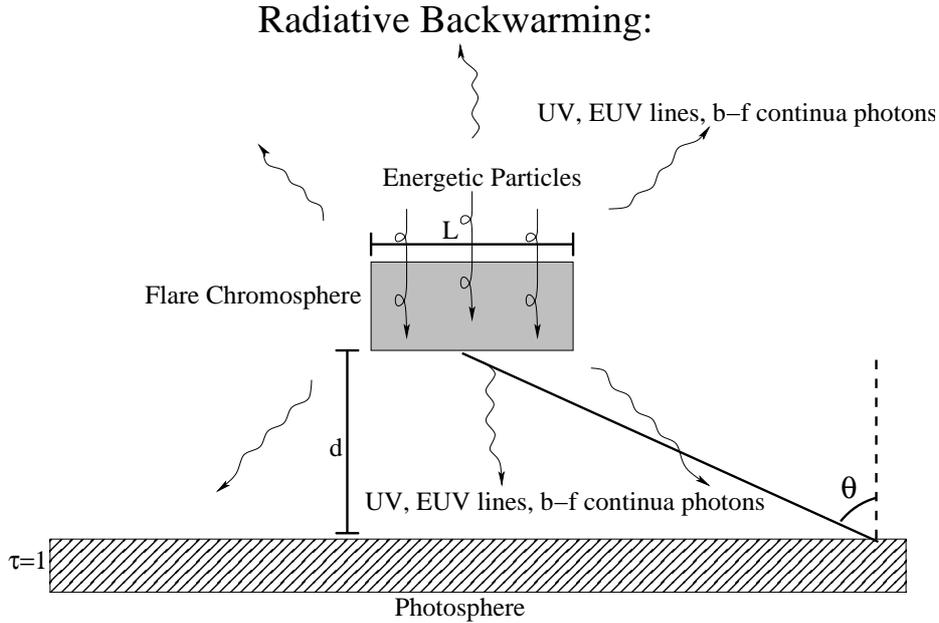}
              }
   \caption{
   Schematic illustration of radiative backwarming in a solar flare.
   Energetic electrons are stopped collisionally in the upper flare 
   chromosphere, raising the temperature of the plasma there.  The
   increased energy input is balanced by an increased radiative output
   in the form of EUV and UV line radiation, and the emission of free--bound
   continua from H and other ions.
   This radiation is emitted in all directions,
   but a signficant fraction of it is re-absorbed in optically thick layers
   near the solar photosphere.  These layers respond with an
   increase of temperature and pressure, with an amplitude that will
   depend sensitively on the energy flux, area coverage, and timing of 
   the impinging radiation.  The emitting layer
   is assumed to be at a height $d$ above the photosphere.  The shape
   of the emitting layer seen from directly below is assumed to be a circular,
   with diameter $L$.
   At an arbitrary location on the photospheric surface, the angle
   between the direction to the center of the source and the vertical direction
   is $\theta$, and the corresponding
   direction cosine $\mu = \cos(\theta)$.  
   }
   \label{figure:backwarming}
   \end{figure}

Observations of the spatial and temporal variation of optical continuum
(white light) emission and hard X-ray emission during solar flares show an
intimate temporal, spatial, and energetic relationship between 
the presence of energetic electrons in the
flare chromosphere and white-light emission from the solar photosphere 
(\opencite{Hudson1992}; \opencite{Metcalf2003}; 
\opencite{Chen2005}; \opencite{Watanabe2010}).  
One 
possible component mechanism of
this connection is
radiative backwarming of the continuum-emitting layers by UV and EUV line and
free--bound emission that is excited by energetic electrons penetrating into
the flare chromosphere, at some distance above the photosphere.  
Since the radiative cooling time of the flare 
chromosphere immediately below the regions undergoing chromospheric evaporation
is so short (\opencite{Fisher1985c}), the temporal variation of UV and EUV 
line emission from plasma within the $10^4-10^5\mathrm{K}$ temperature range 
will closely track
heating by energetic electrons, detected as
hard X-ray emission emitted from footpoints in the flare chromosphere.
The backwarming
scenario is illustrated schematically in Figure \ref{figure:backwarming}.

Estimates of the continuum opacity and atmospheric density near the solar
photosphere indicate that the layer responsible for most of 
the optical continuum emission is about one continuum
photon mean-free path thick, or
roughly 70 km.  Thus 
most of
the energy from the impinging 
backwarming radiation will be reprocessed into optical continuum emission 
within a thin layer near the solar photosphere.

Does the absorption of this radiation within this thin layer result in a
significant downward force, via a pressure perturbation from enhanced heating?  
This mechanism has
been suggested by \inlinecite{Donea2006}, \inlinecite{Moradi2007}, 
and \inlinecite{Lindsey2008} 
as a potential source for ``sunquake'' 
acoustic emission seen during a few solar flares.  
Here, we compare and contrast this mechanism
of creating a force perturbation
with that from the Lorentz force change described earlier.

The simplest estimate of the pressure change
is to assume that the backheated photospheric plasma is frozen
in place during the heating process, and that its temperature will rise to
a level where the black-body radiated energy flux
equals the combined output of the preflare
solar radiative flux plus the incoming flare energy flux due to backwarming.

What is the flux of energy from backwarming available to heat the photosphere?
In order to compute this flux, we must first estimate the fraction
of the non-thermal electron energy flux that is balanced by
chromospheric UV/EUV line and free--bound continuum
emission, and the fraction of this radiated
energy that impinges on the nearby solar photosphere.  We can use the
estimate presented in Section \ref{section:gasbag} and plotted in Figure
\ref{figure:fratioplot} for 
$F_\mathrm{evap} / F_\mathrm{nte}$ to find the flux of energy $F_\mathrm{rad}$
that is converted from non-thermal electrons into radiated energy,
\be
F_\mathrm{rad} \simeq (1 - F_\mathrm{evap}/F_\mathrm{nte})\ F_\mathrm{nte}\ .
\label{equation:frad}
\ee
Not all of this energy flux will be available for backwarming,
because the resulting radiation is emitted isotropically, while the photosphere
lies beneath the radiating source.  If the lateral dimension
$L$ of the illuminating region is much larger than the distance $d$
above the photosphere, then up to half the radiated energy flux will irradiate 
the photosphere
directly beneath the source (see Figure \ref{figure:backwarming}).  If
the ratio $d/L$ is of order unity, on the other hand,
there is a substantially reduced geometrical
dilution factor [$f_\mathrm{geom}$] that must multiply $F_\mathrm{rad}$  
to determine the flux of radiated energy that is incident on the photosphere
beneath the source.  We estimate the geometrical dilution factor
$f_\mathrm{geom}$ as
\be
f_\mathrm{geom} = {1 \over 2} \left( 1- {2d / L \over \sqrt{1 + 4 (d / L)^2}}
\right) \mu^3\ ,
\ee
where for simplicity this expression 
assumes that the shape of the irradiating source 
shown in Figure \ref{figure:backwarming} is
a circular disk of diameter $L$.  If the position of interest on
the photosphere is not directly beneath the illuminating
source, but instead is located at
an angle $\theta$ away from the vertical direction (see Figure 
\ref{figure:backwarming}), this expression
includes a factor of
$\mu^3$, where $\mu = \cos(\theta)$, accounting for both increased distance
from the source to the given point on the photosphere and the oblique 
angle of the irradiating source relative to the normal direction.
With the geometrical dilution factor determined, this 
results in the following estimate of the elevated photospheric temperature [$T$]:
\be
\sigma T^4 = \sigma T_{0}^4 + f_\mathrm{geom} F_\mathrm{rad}\ ,
\label{equation:tblackbody}
\ee
where $T_{0}$ is the non-flare photospheric effective temperature.  This
expression can be re-written as
\be
{\Delta T \over T_0} = \left( 1+ {f_\mathrm{geom} F_\mathrm{rad} \over 
\sigma T_0^4} \right) ^{1/4} - 1\ ,
\label{equation:deltatovert}
\ee
where $\Delta T / T_0$ is the ratio of the temperature rise to the 
preflare photospheric temperature.

\begin{figure}    
   \centerline{\includegraphics[width=1.0\textwidth,clip=]{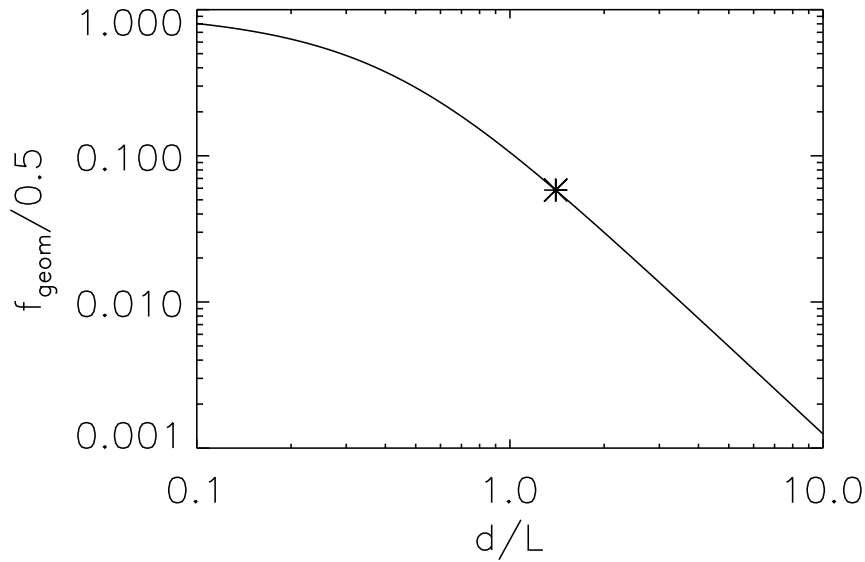}
              }
   \caption{
   Computed ratio of geometrical dilution factor to the plane-parallel value
   of $1/2$ as a function of $d / L$, where $d$ is the height of the source
   above the photosphere, and $L$ is the diameter of the source, which is
   assumed to have a circular shape.  The asterisk on the curve
   shows the value for
   $d/L = 1.4$ (see text).  The plot assumes $\mu=1$, $i.e.$ for a point
   on the photosphere directly beneath the irradiating source.
   }
   \label{figure:geomplot}
   \end{figure}

\begin{figure}    
   \centerline{\includegraphics[width=1.0\textwidth,clip=]{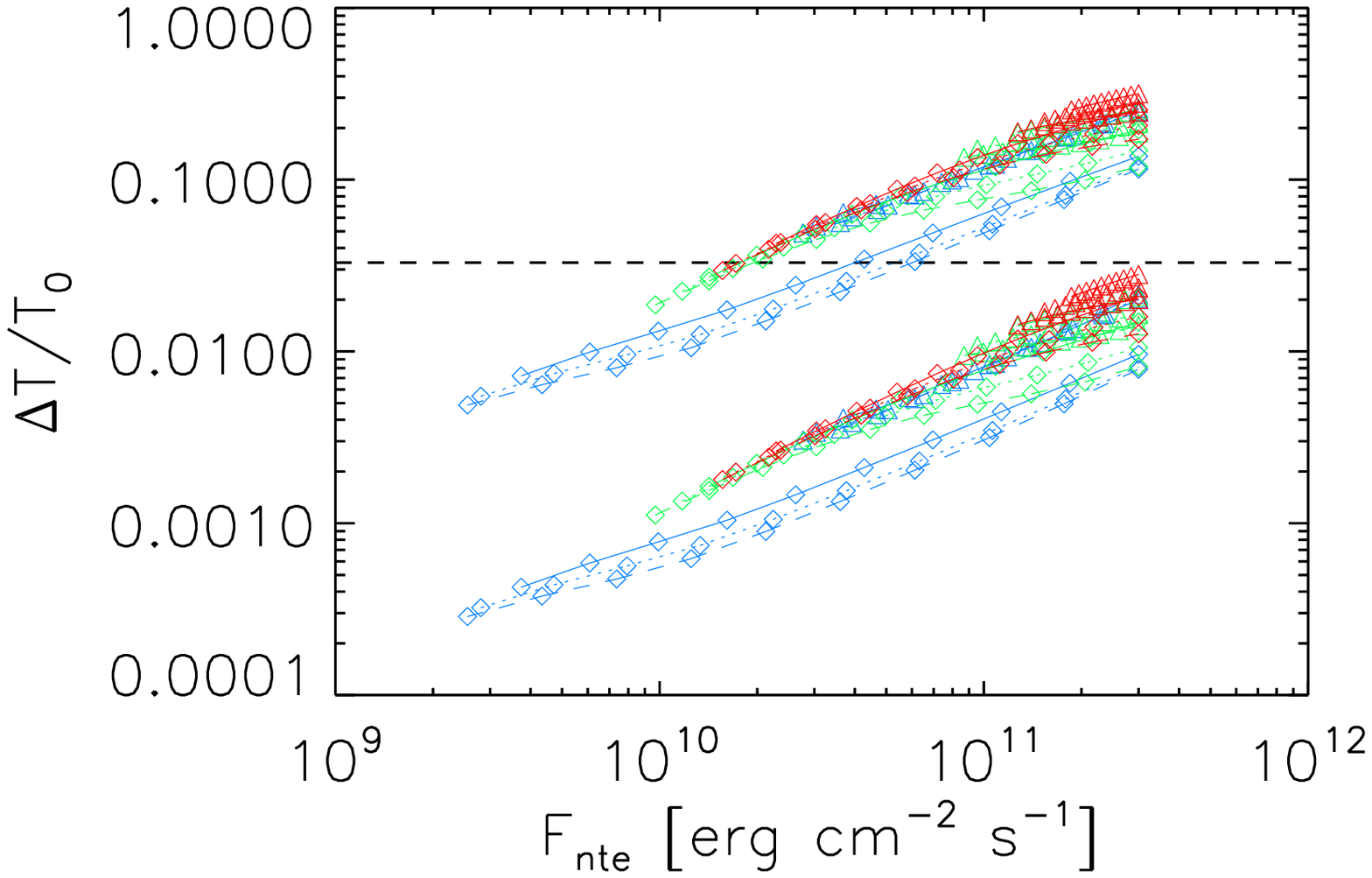}
              }
   \caption{
   Computed ratio of temperature rise to background photospheric temperature
   as a function of the flux of energy in non-thermal electrons, using
   Equation (\ref{equation:deltatovert}).  
   Diamonds indicate a pre-flare coronal pressure of
   0.3 ${\rm dyne}\ {\rm cm}^{-2}$, 
   while triangles indicate a higher
   pre-flare coronal pressure of 3.0 ${\rm dyne}\ {\rm cm}^{-2}$.
   Blue symbols indicate a 10 KeV low-energy cutoff, green indicates
   a 20 KeV low-energy cutoff, and red indicates a
   25 KeV low-energy cutoff.  
   Solid curves indicate an electron spectral index
   $\delta = 4$, dotted curves $\delta = 5$, and dashed
   curves $\delta = 6$.  The geometrical dilution factor in the
   upper set of curves was set to $1/2$, consistent with plane-parallel
   geometry, in which the horizontal dimension $L$ of the illuminating region
   is much greater than the height $d$ of the illuminating region above the
   photosphere.  The lower set of curves was computed by setting $d/L$ to
   $1.4$, consistent with a 1000\ km diameter 
   source illuminating the photosphere from a height of 1400\ km (see text).
   The dashed horizontal line indicates the temperature ratio needed to
   match a vertical Lorentz force surface density of
   $2500\ \mathrm{dyne\ cm}^{-2}$.
   }
   \label{figure:tratio}
   \end{figure}

Next, we
estimate the height above the photosphere for the source of the
backwarming radiation.  To do this we find the change in depth between the
preflare and flare transition region, using the same explosive evaporation
model described in Section \ref{section:gasbag}.  For cases where the assumed
flux in non-thermal electrons exceeds 
$10^{11}$ $\mathrm{erg\ cm}^{-2}\mathrm{\ s}^{-1}$, the 
depth of the flare transition region relative to the preflare
transition region moves downwards by distances ranging from $\approx$\,100 km for the
dense preflare corona, up to $\approx$\,600 km for the tenuous preflare corona.  The
primary source of the backwarming radiation will be in the layers immediately
below the flare transition region.  Assuming an approximate distance between
the photosphere and preflare transition region of 2000 km ($e.g.$ model F
of \opencite{Vernazza1981}), thus leads to an expected distance that ranges
from 1400 to 1900 km between the backwarming source and the photosphere.
We then estimate
the geometrical dilution of a UV/EUV emitting flare kernel with roughly
$L \approx 1000\ \mathrm{km}$ 
in horizontal extent (based on estimated flare kernel areas of roughly 
one arcsecond$^2$) located in the flare chromosphere
at a distance $d=1400\ \mathrm{km}$ above the photosphere, 
in keeping with the above distance
estimates, and find a geometrical dilution factor assuming $d/L = 1.4$ of 
$f_\mathrm{geom} = 0.029$, shown as the asterisk in Figure
\ref{figure:geomplot}.  The low value of $f_\mathrm{geom}$
stems from the fact that the source as seen from the photosphere 
subtends a solid angle of less than 0.4 steradians,
compared with the $4 \pi$ steradians over which the radiation is emitted.

In Figure \ref{figure:tratio}, we use Equation (\ref{equation:deltatovert})
to plot the temperature enhancement as
a function of non-thermal electron energy flux for two different assumed
values 
of $f_\mathrm{geom}$, $1/2$, corresponding to a widespread (plane-parallel)
source of backwarming radiation, and $0.029$, corresponding to $d/L = 1.4$.
This Figure shows a wide range in possible values of $\Delta T / T$ for
a commonly assumed range of non-thermal electron energy fluxes.  This
fact, coupled with the wide range of possible values for $f_\mathrm{geom}$,
illustrates the difficulty in making broad conclusions about the
effectiveness of backwarming in perturbing the photosphere.  

The horizontal line in Figure \ref{figure:tratio} corresponds to the
temperature ratio that would lead to a pressure increase comparable to the
estimated Lorentz force surface density of $2500\ \mathrm{dyne\ cm^{-2}}$
for the large flare discussed in \inlinecite{Hudson2008}.  Here, we
adopt a pre-flare
photospheric pressure of $7.6 \times 10^4\ {\rm\ dyne\ cm}^{-2}$ (see model S
of \opencite{Dalsgaard1996}).  In the limit of
large-scale source size ($d \ll L$), $f_\mathrm{geom} = 1/2$, and there is
a wide range of non-thermal electron energy fluxes which yield pressure
increases which could be comparable or even greater than the above Lorentz
force example.  On the other hand, for a small 1000\ km flare kernel size,
we find temperature enhancements that are comparable to the candidate
Lorentz force value only for the very largest non-thermal electron 
energy fluxes we have considered.
Regarding the non-thermal electron energy flux levels, we must point out that
recent RHESSI and \textit{Hinode} observations (\opencite{Krucker2011}) of the white-light flare of 6 December 2006 indicate a value of $F_\mathrm{nte}$ of
$10^{12}-10^{13}\ \mathrm{erg\ cm}^{-2}\mathrm{\ s}^{-1}$, a value that
greatly exceeds our assumed energy flux range in Figure \ref{figure:tratio},
and which would result in a pressure increase that greatly exceeds the Lorentz
force estimate in that figure, even for a small value of $L/d$.  We also
point out, on the other hand, an observed sunquake (15 February 2011), for
which no white-light enhancement was observed at all (C.A. Lindsey and J.C. Mart\'inez-Oliveras, 2011, Private communication),
indicating that in this case backwarming did not play a signficant role.

We must caution that our perturbed pressure estimates for backwarming
are probably overestimates.
Our treatment assumes the temperature changes instantaneously 
(ignoring the
time-lag due to the finite heat capacity of the photospheric plasma), 
and it assumes the photospheric plasma
is frozen in place and
does not respond dynamically to the increased heating (the plasma 
should expand in response to the enhanced heating on a sound-crossing
time -- for a 70 km thick photospheric layer, with 
$C_s \approx 8 {\rm\ km\ s}^{-1}$, this is $\approx$\,ten seconds).
This treatment also ignores the possibility of
multi-step radiative reprocessing, in which the backwarming radiation that
reaches the photosphere comes not from the primary source in the flare
chromosphere, but from secondary backheating sources, where the UV line
emission is first converted via backwarming to other radiation mechanisms
($e.g.$ H bound--free continuum emission),
before finally reaching the photosphere.  
Each step of a multi-step
reprocessing will result in further dilution of the energy flux reaching
the photosphere.

In summary, our simple estimate of backwarming-induced temperature and
pressure increases shows a wide range of possible outcomes.  To be competitive
with the Lorentz-force surface density taken from \inlinecite{Hudson2008},
requires either backwarming sources that are much wider than their height
above the photosphere, or for small flare kernel sizes, energy fluxes well
in excess of $10^{11}\ \mathrm{erg\ cm^{-2}\ s^{-1}}$.  Our pressure 
enhancement estimates are probably upper limits, since they ignore heat
capacity effects, expansion of the heated photospheric plasma, and any
secondary reprocessing.  Nevertheless these estimates provide useful guidelines
for future, more detailed investigations of flare-driven backwarming.

\section{Conclusions}
\label{section:conclusions}

We derive an expression for the vertical and horizontal components of the
Lorentz-force change
implied by observed magnetic-field changes occurring over the course of a
solar flare.  The Lorentz-force change acting on the
outer solar atmosphere,
Equations (\ref{equation:deltafr})\,--\,(\ref{equation:deltafh}), is
balanced exactly by a corresponding Lorentz force change  acting 
on the photosphere and below: Equations 
(\ref{equation:deltafrdown})\,--\,(\ref{equation:deltafhdown}).
The Lorentz force change, integrated over the time period over which the
change occurs, defines an impulse.  The impulse defines a momentum
increase, given
in Equations (\ref{equation:impulser}) and (\ref{equation:impulseh}).
The radial component of the impulse, acting on the outer solar
atmosphere, is then used to derive an upper limit to the mass of CME ejecta
that escape from the Sun: Equation (\ref{equation:massul}).

We show that our expression for the vertical
Lorentz-force change acting on the
solar interior, Equation (\ref{equation:deltafrdown}),
generalizes our earlier result in \inlinecite{Hudson2008},
in that it includes horizontal as well as vertical (radial) forces.  It is also
more accurate, in that it
does not assume a first-order expansion of changes in the magnetic
field.

The balance between the Lorentz forces acting on the solar atmosphere and
the solar interior leads us to suggest a possible
connection between the upward momentum
in flare ejecta and the downward momentum in the solar interior, and leads to
the possibility of using helioseismic measurements of ``sunquakes'' to study
the properties of eruptive flares and CMEs.

For the purpose of further elucidating the 
physical origins of ``sunquake'' acoustic emission,
we also estimate force perturbations in the photosphere due to changes in
gas pressure driven by chromospheric evaporation and from
radiative backwarming of the photosphere during solar flares.  We 
find, for chromospheric evaporation in flares, an upper limit
of $\approx 2000\ {\rm\ dyne\ cm}^{-2}$ 
for a pressure increase in the upper chromosphere.
We show that this pressure increase will lead to the downward propagation
of a chromospheric condensation, (a dense region behind a downward moving
shock-like disturbance), but the chromospheric condensation can propagate
to column depths of at most a few percent of the photospheric depth; the
subsequent propagation to the photosphere is by means of 
an acoustic disturbance.  Whether this
acoustic disturbance is significant at photospheric depths, when compared
to the Lorentz force per unit area, is not yet clear, and will require
a more detailed analysis of acoustic-wave propagation and dissipation effects
in the solar chromosphere.  We compare the Lorentz force 
to gas-pressure changes driven by radiative backwarming, and find the
latter mechanism could be comparable to, or greater than, the Lorentz force
if the region being
energized by flare non-thermal electron heating has a horizontal 
extent much
greater than its height above the photosphere, or for smaller heated regions,
if the nonthermal electron energy flux greatly exceeds
$10^{11}\ \mathrm{erg\ cm^{-2}\ s^{-1}}$.  We caution that our estimates
of pressure changes due to backheating are probably overestimates.

To summarize, the primary source for energy release in eruptive solar
flares is most likely the solar magnetic field in strong-field, low-$\beta$
active regions.  It then makes sense that changes in the magnetic field itself
will have a more direct and larger impact on the atmosphere than changes
that are due to secondary flare processes, such as the production of
energetic particles, gasdynamic motions, and enhanced radiative output,
all of which are assumed to be driven ultimately by the release of magnetic
energy.

\begin{acks}
This work was supported by the NASA Heliophysics Theory Program (grants
NASA-NNX08AI56G and NASA-NNX11AJ65G), 
by the NASA LWS TR\&T program (grants NNX08AQ30G and NNX11AQ56G), by
the NSF SHINE program (grants ATM0551084 and ATM-0752597), 
by the NSF
core AGS program (grant ATM-0641303) funding our efforts in support of the
University of Michigan's CCHM project, 
NSF NSWP grant AGS-1024682, and NSF core grant AGS-1048318.  HSH
acknowledges support from NASA Contract NAS5-98033.  We thank the US
taxpayers for making this research possible.  We wish to acknowledge the
role that Dick Canfield has played in realizing the importance of global
momentum balance in the dynamics of solar-flare plasmas, through several
seminal papers on this topic in the 1980s and 1990s.
We thank the referee, Charlie Lindsey, for noting an error in our initial
discussion of the depth dependence of the gas pressure perturbation for
downward propagating acoustic waves.  We also thank him for providing 
an exceptionally thoughtful, thorough, and detailed
critique of this article, which improved it greatly.
\end{acks}

   


\end{article} 

\end{document}